\documentclass[%
reprint,
 amsmath,amssymb,
 aps,
]{revtex4-1}

\usepackage[left]{lineno}

\usepackage{graphicx}

\usepackage{dcolumn}

\usepackage{bm}

\usepackage{float}

\usepackage{diagbox}

\begin{document}

\preprint{APS/123-QED}

\title{Beam energy dependence of net-$\Lambda$ fluctuations measured by the STAR experiment at RHIC}



\author{
J.~Adam$^{6}$,
L.~Adamczyk$^{2}$,
J.~R.~Adams$^{39}$,
J.~K.~Adkins$^{30}$,
G.~Agakishiev$^{28}$,
M.~M.~Aggarwal$^{40}$,
Z.~Ahammed$^{59}$,
I.~Alekseev$^{3,35}$,
D.~M.~Anderson$^{53}$,
R.~Aoyama$^{56}$
% ...
\\
(STAR Collaboration)
\vspace{2mm}
}

% ...
\address{$^{2}$AGH University of Science and Technology, FPACS, Cracow 30-059, Poland}
\address{$^{3}$Alikhanov Institute for Theoretical and Experimental Physics, Moscow 117218, Russia}
% ...
\address{$^{6}$Brookhaven National Laboratory, Upton, New York 11973}
% ...
\address{$^{28}$Joint Institute for Nuclear Research, Dubna 141 980, Russia}
% ...
\address{$^{30}$University of Kentucky, Lexington, Kentucky 40506-0055}
% ...
\address{$^{35}$National Research Nuclear University MEPhI, Moscow 115409, Russia}
% ...
\address{$^{39}$Ohio State University, Columbus, Ohio 43210}
\address{$^{40}$Panjab University, Chandigarh 160014, India}
% ...
\address{$^{53}$Texas A\&M University, College Station, Texas 77843}
% ...
\address{$^{56}$University of Tsukuba, Tsukuba, Ibaraki 305-8571, Japan}
% ...
\address{$^{59}$Variable Energy Cyclotron Centre, Kolkata 700064, India}
% ...


\begin{abstract}
The measurements of particle multiplicity distributions have generated considerable interest in understanding the fluctuations of conserved quantum numbers in the Quantum Chromodynamics (QCD) hadronization regime, in particular near a possible critical point and near the chemical freeze-out. Net-protons and net-kaons have been used as proxies for the net-baryon number and net-strangeness, respectively. We report the measurement of efficiency and centrality bin width corrected cumulant ratios ($C_{2}/C_{1}$, $C_{3}/C_{2}$) of net-$\Lambda$ distributions, in the context of both strangeness and baryon number conservation, as a function of collision energy, centrality and rapidity. The results are for Au + Au collisions at five beam energies ($\sqrt{s_{NN}}$ = 19.6, 27, 39, 62.4 and 200 GeV) recorded with the Solenoidal Tracker at RHIC (STAR). We compare our results to the Poisson and negative binomial (NBD) expectations, as well as to Ultra-relativistic Quantum Molecular Dynamics (UrQMD) and Hadron Resonance Gas (HRG) model predictions. Both NBD and Poisson baselines agree with data within the statistical and systematic uncertainties. UrQMD describes the measured net-$\Lambda$ $C_{1}$ and $C_{3}$ at 200 GeV reasonably well, but deviates from $C_{2}$, and the deviation increases as a function of collision energy. The ratios of the measured cumulants show no features of critical fluctuations. The chemical freeze-out temperatures extracted from a recent HRG calculation, which was successfully used to describe the net-proton, net-kaon and net-charge data, indicate $\Lambda$ freeze-out conditions similar to those of kaons. However, large deviations are found when comparing to temperatures obtained from net-proton fluctuations. The net-$\Lambda$ cumulants show a weak, but finite, dependence on the rapidity coverage in the acceptance of the detector, which can be attributed to quantum number conservation.  

\begin{description}
\item[PACS numbers]25.75.-q

\end{description}

\end{abstract}


\maketitle

\section{\label{sec:level1}Introduction}

Relativistic heavy-ion collisions provide significant information on the nuclear matter phase transition under extreme temperatures and energy densities. The Beam Energy Scan (BES) program at the Relativistic Heavy Ion Collider (RHIC) was established for the purpose of studying the QCD phase diagram as a function of temperature ($T$) and baryon chemical potential ($\mu_{B}$). Lattice QCD calculations suggest that the phase transition from Quark Gluon Plasma (QGP) to hadron gas at low $\mu_{B}$ is a smooth cross-over \cite{Aoki:2006we}, while at relatively high $\mu_{B}$ effective chiral theories predict a first order transition \cite{Ejiri:2008xt,Alford:1997zt}. Probing the existence of an end point for the first order phase transition, i.e. the QCD critical point, and mapping the chemical freeze-out process of hadrons at different $T$ and $\mu_{B}$ are two major goals of the fluctuation measurements in the BES program at RHIC.

\par

Fluctuations of conserved quantum numbers, in particular charge ($Q$), baryon number ($B$) and strangeness ($s$), show sensitivity to the QCD phase transition \cite{Stephanov:2008,Stephanov:2011,Gupta:2011}. These quantum numbers can be represented by experimentally measured net-particle multiplicities \cite{Luo:2017faz}. Potential proxies for the net-charge ($\Delta Q$), net-baryon number ($\Delta B$) and the net-strangeness ($\Delta s$) are the net-charged particle, net-proton and the net-kaon multiplicities, respectively. Each proxy has unique caveats in fully reflecting the behavior of the conserved quantum number during the transition. Thus, in this paper we exclusively use comparisons to phenomenological approaches, such as Ultra-relativistic Quantum Molecular Dynamics (UrQMD) and Hadron Resonance Gas (HRG), which can model these caveats when describing the experimental results, rather than compare directly to lattice QCD calculations. Moments of net-particle multiplicity distributions such as mean ($M$), standard deviation ($\sigma$), skewness ($S$) and kurtosis ($\kappa$) are related to the thermodynamic susceptibilities, $\chi^{(n)}_{i}$, where $n$ is the order of the susceptibility and $i$ stands for the type of conserved quantum number. These susceptibilities can also be written in terms of cumulants, $C_{n}$ as: $\chi^{(n)}_{i}$ $=$ $(1/VT^{3})C_{n}$, where $V$ and $T$ stand for volume and temperature, respectively \cite{Ratti:2018ksb}. Moment products or ratios of net-particle multiplicity distributions are related to the volume independent susceptibility ratios as: $\sigma^{2}/M$ $=$ $\chi^{(2)}_{i}/\chi^{(1)}_{i}$ and $S\sigma$ $=$ $\chi^{(3)}_{i}/\chi^{(2)}_{i}$. Thermodynamic susceptibilities have been modeled in lattice QCD at $\mu_{B}$ $=$ 0 \cite{Ratti:2018ksb,Gavai:2016,Gavai:2008,Bazavov:2012-1,Bazavov:2012-2,Borsanyi:2013,Alba:2014, Karsch:2015} and in HRG models at finite $\mu_{B}$ \cite{KARSCH:2011,Garg:2013a,FU:2013,Nahrgang:2015} as a function of $T$ and $\mu_{B}$. By comparing the experimentally measured fluctuations of net-particle multiplicity distributions with the theoretically calculated quantum number susceptibilities at different collision energies, the freeze-out temperature of the strongly interacting matter can be determined for different chemical potentials. In this context, the net-kaon fluctuation measurements have been studied as a proxy for net-strangeness and the net-proton results were used to study net-baryon number \cite{Net_K_STAR,Adamczyk:2013dal}. The freeze-out temperatures extracted by comparing the latest HRG model calculations with STAR net-kaon fluctuation measurements revealed higher values than the freeze-out temperatures extracted from previous measurements by STAR of the net-charge/proton fluctuations \cite{Bellwied:2018tkc}.

\par

The $\Lambda$ carries both baryon and strangeness quantum number. Thus, the study of the event-by-event net-$\Lambda$ ($\Lambda$ multiplicity  $-$ $\overline{\Lambda}$ multiplicity) fluctuations is important for the understanding of the freeze-out temperature in the context of both baryon number and strangeness conservation. Predicted sequential hadronization \cite{Bellwied_FO_sequential} can be addressed by comparing the net-$\Lambda$ fluctuations with the latest HRG results \cite{Bellwied:2018tkc}, which calculate the cumulant ratios using the freeze-out conditions from different fluctuation measurements of net-quantum numbers. Additionally, net-$\Lambda$'s paired with net-kaons provide a more complete measurement of the net-strangeness fluctuations, while paired with net-protons the results are closer to a complete measurement of the fluctuations of the net-baryon number. Net-$\Lambda$ fluctuations are also the main contribution to the measurement of the off-diagonal baryon-strangeness correlator, and thus add to recent STAR and theoretical publications on this issue \cite{AAA, Bellwied:2019pxh, BBB}.

\par

In this paper, the first measurements of the net-$\Lambda$ fluctuations in Au+Au collisions using the STAR (Solenoidal Tracker At RHIC) experiment are presented as a function of collision energy, centrality and rapidity. Results are compared to the Poisson and Negative Binomial Distribution (NBD) baselines, as well as to the UrQMD predictions \cite{UrQMD_Cite} and the predictions from the latest HRG model \cite{Bellwied:2018tkc}.

\section{\label{sec:level2}Analysis details}

\par

For this analysis, we use $N$ to represent the measured event-by-event observable, which is the event-by-event net-$\Lambda$ multiplicity ($\Delta N_{\Lambda}$ = $N_{\Lambda}$ - $N_{\overline{\Lambda}}$). The average value of the observable $N$ is represented by $\langle N \rangle$. The deviation of $N$ from the mean is given by $\delta N$ = $N$ - $\langle N \rangle$. The first three cumulants ($C_{1}$, $C_{2}$ and $C_{3}$,) of the event-by-event distribution of $N$ can be written as

\begin{equation}
C_{1} = \langle N \rangle,
\end{equation}
\vspace{-0.9cm}
\begin{equation}
C_{2} = \langle (\delta N)^{2} \rangle,
\end{equation}
\vspace{-0.9cm}
\begin{equation}
C_{3} = \langle (\delta N)^{3} \rangle.
\end{equation}

\noindent{The mean, variance and skewness of the distribution are related to the cumulants through}

\begin{equation}
M = C_{1}, \;  \sigma^{2} = C_{2}, \; S = \frac{C_{3}}{(C_{2})^{\frac{3}{2}}}. 
\end{equation}

\noindent{The products/ratios presented in this analysis are}

\begin{equation}
\frac{\sigma^{2}}{M} = \frac{C_{2}}{C_{1}}, \; S\sigma = \frac{C_{3}}{C_{2}}.
\end{equation}

\par

The event-by-event $\Lambda$ and $\overline{\Lambda}$ multiplicities were measured for Au+Au minimum bias events at $\sqrt{s_{NN}}$ = 19.6, 27, 39, 62.4 and 200 GeV collision energies. The 39 and 62.4 GeV data were collected in 2010; all other energies were recorded in 2011. The number of events analyzed were: $16\times10^{6}$, $33\times10^{6}$, $77\times10^{6}$, $27\times10^{6}$ and $199\times10^{6}$ for $\sqrt{s_{NN}}$ = 19.6, 27, 39, 62.4 and 200 GeV, respectively. Although the previously analyzed identified net-particle distributions in STAR for protons and kaons reached down to $\sqrt{s_{NN}}$ = 7.7 GeV, the available statistics for net-$\Lambda$'s, in particular for the higher moments, were too small to extend the analysis below $\sqrt{s_{NN}}$ = 19.6 GeV. We hope to extend the analysis to the lower energies, even for the net-$\Lambda$'s higher moments, with the datasets taken during the second beam Energy Scan campaign at RHIC. For the precise determination of the primary vertex (PV), the STAR Time Projection Chamber (TPC) and the Vertex Position Detectors (VPDs) were used. Only the collisions occurring within a distance of 30 cm from the center of the detector along the beam line were chosen. Effects from possible interactions of the beam with the beam pipe were minimized by rejecting collisions with a radial distance of 2 cm or greater from the center of the detector in the transverse plane. Pile-up events were removed by only selecting collisions with a less than 2 cm difference between the PV measurements along the beam line obtained from the TPC and VPD. 

\par

The reconstruction of $\Lambda$ ($\overline{\Lambda}$) baryons was restricted to the decay channel: $\Lambda$ ($\overline{\Lambda}$) $\rightarrow$ $p$($\overline{p}$) + $\pi^{-}$($\pi^{+}$), which has a branching ratio of 63.9\% $\pm$ 0.5\%. The STAR TPC was used as the main tracking and charged daughter particle identification device in this analysis \cite{Anderson:STAR_TPC:2003}. Ionization energy loss per unit length ($dE/dx$) of a charged particle $X$ in the TPC gas was used to calculate the quantity $n\sigma_{X}$ which is defined as

\begin{equation}
    n\sigma_{X} = \frac{\ln[(dE/dx)_{\rm measured}/(dE/dx)_{\rm theory}]}{\sigma_{X}},
\end{equation}

\noindent{where $(dE/dx)_{measured}$ is the measured ionization energy loss from TPC, $(dE/dx)_{theory}$ is the theoretical expectation of the ionization energy loss from the Bichsel formula \cite{Bichsel:2006:Bethe} and $\sigma_{X}$ is the $dE/dx$ resolution of the TPC. The identification of species $X$ ($p$ ($\overline{p}$) or $\pi^{+}$ ($\pi^{-}$)) was done by imposing a cut, $|$$n\sigma_{X}$$|$ $<$ $2.0$ within a rapidity coverage of $|y|$ $<$ $1.0$ and a transverse momentum of $p_{T}$ $>$ $0.05$ GeV/c. The invariant mass ($M_{\rm inv}$) of $\Lambda$ ($\overline{\Lambda}$) was reconstructed using the energy, momentum and rest-mass of the daughter particles, $p$ ($\overline{p}$) and $\pi^{+}$ ($\pi^{-}$). 

The selection of possible $\Lambda$ and $\overline{\Lambda}$ ($V^{0}$) candidates was done by imposing an invariant mass cut, $1.11$ $<$ $M_{\rm inv}$ (GeV/c$^{2}$) $<$ $1.12$ and applying topological cuts as shown in Table I. These tight cuts were applied in order to achieve a signal purity of greater than $90\%$ for all collision energies. Figure 1 shows typical $\Lambda$ and $\bar\Lambda$ invariant mass plots. No further background subtraction was applied, since studies, based on an analysis of the cumulants in the invariant mass side-bands, showed that the remaining contamination contributes in a negligible way to the systematic error and to the absolute values of the cumulants themselves. A detailed description of particle reconstruction, track quality, decay vertex topology cuts and calculation of the detection efficiency can be found in Ref. \cite{Ackermann:2000:reconstruct-1,Adler:2002:reconstruct-2}.}

\begin{table}[hbt]
    \caption{Topological cuts used for the extraction of $V^{0}$s event-by-event (DCA: Distance of Closest Approach, PV: Primary Vertex).}
    \medskip
    \centering
    \begin{tabular}{c | c}
    \hline\hline
    Topological Parameter & Cut \\
    \hline\hline
         DCA of $p$ ($\overline{p}$) to PV & $>$ 0.5 cm  \\
         DCA of $\pi^{-}$ ($\pi^{+}$) to PV & $>$ 1.5 cm \\
         DCA of $p$ ($\overline{p}$) to $\pi^{-}$ ($\pi^{+}$) & $<$ 0.6 cm \\
         DCA of $V^{0}$ to PV & $<$ 0.5 cm \\
         $V^{0}$ decay length & $>$ 3.0 cm \\
    \hline\hline
    \end{tabular}
    \label{tab:my_label_01}
\end{table}

Only $\Lambda$'s and $\overline{\Lambda}$'s produced in a rapidity window $|y|$ $<$ $0.5$, and with a transverse momentum within $0.9$ $<$ $p_{T}$(GeV/c) $<$ $2.0$ were used in this analysis. The low cut-off is driven by the falling reconstruction efficiency; the high cut-off is determined by the choice of using the for particle identification only the TPC. No feed-down correction from multi-strange baryon decays is applied, since we found that although the feed-down impacts the single cumulants shown in Fig.2 its impact on the cumulants ratios in Figs.3-9 is negligible. The calculation of reconstruction efficiency was based on the probability of finding Monte Carlo generated particles after passing them through a TPC detector response simulation and then embedding them into real events prior to reconstruction. As an example, Table II shows the resulting $p_{T}$-averaged efficiencies for $\Lambda$'s for the bins of highest collision centrality as a function of the beam energy. Within a chosen rapidity window, the particle reconstruction efficiency depends on the transverse momentum. The net-$\Lambda$ cumulants were corrected for the reconstruction efficiency following the method in Ref. \cite{Nonaka:2017:EffCorr-pT-dep}, which takes the $p_{T}$ dependence of the reconstruction efficiency into account. The number of $p_{T}$ bins was varied from 3 to 6 bins in the aforementioned range without any discernible effect on the correction factors. The average reconstruction efficiency ($\langle \epsilon \rangle$) in each $p_{T}$ bin was calculated as

\begin{figure*}
\centering
\includegraphics[width=0.75\textwidth]{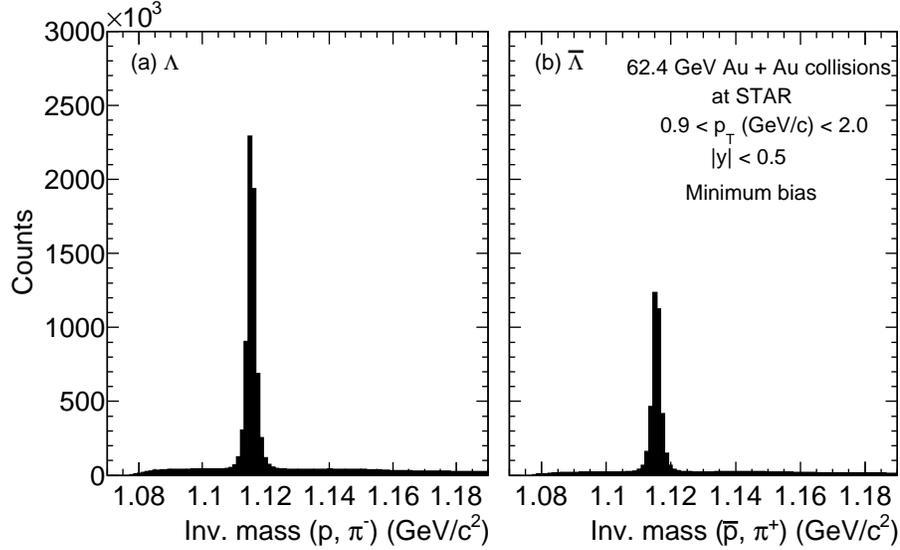}
\caption{\label{F_1}Reconstructed invariant mass distribution of a) $\Lambda$ and b) $\bar{\Lambda}$ for 62.4 GeV Au + Au collisions. Topological cuts shown in Table I are applied.}
\end{figure*}

\begin{table}[]
    \caption{Reconstruction efficiency of $\Lambda$, calculated in the transverse momentum range 0.9 $<$ $p_{T}$ (GeV/c) $<$ 2.0, and in centrality classes, 0-5\% to 20-30\%, using Eq.7 for five beam energies: 19.6, 27, 39, 62.4 and 200 GeV.}
    \medskip
    \centering
    \begin{tabular}{c | c | c | c | c}
    \hline\hline
    \backslashbox{Energy}{Centrality} & 20-30\% & 10-20\% & 5-10\% & 0-5\% \\
    \hline\hline
    19.6 GeV & 0.236 & 0.223 & 0.215 & 0.183 \\
    27 GeV & 0.236 & 0.226 & 0.207 & 0.186 \\
    39 GeV & 0.242 & 0.228 & 0.213 & 0.182 \\
    62.4 GeV & 0.224 & 0.204 & 0.183 & 0.151 \\
    200 GeV & 0.202 & 0.176 & 0.147 & 0.127 \\
    \hline\hline
    \end{tabular}
    \label{tab:my_label_02}
\end{table}

\begin{equation}
    \langle \epsilon \rangle = \frac{\int_{a}^{b}\Big(\frac{dN}{dp_{T}}\Big)_{RC}dp_{T}}{\int_{a}^{b}\Big(\frac{dN}{dp_{T}}\Big)_{MC}dp_{T}},
\end{equation}

\begin{figure*}
\centering
\includegraphics[width=0.85\textwidth]{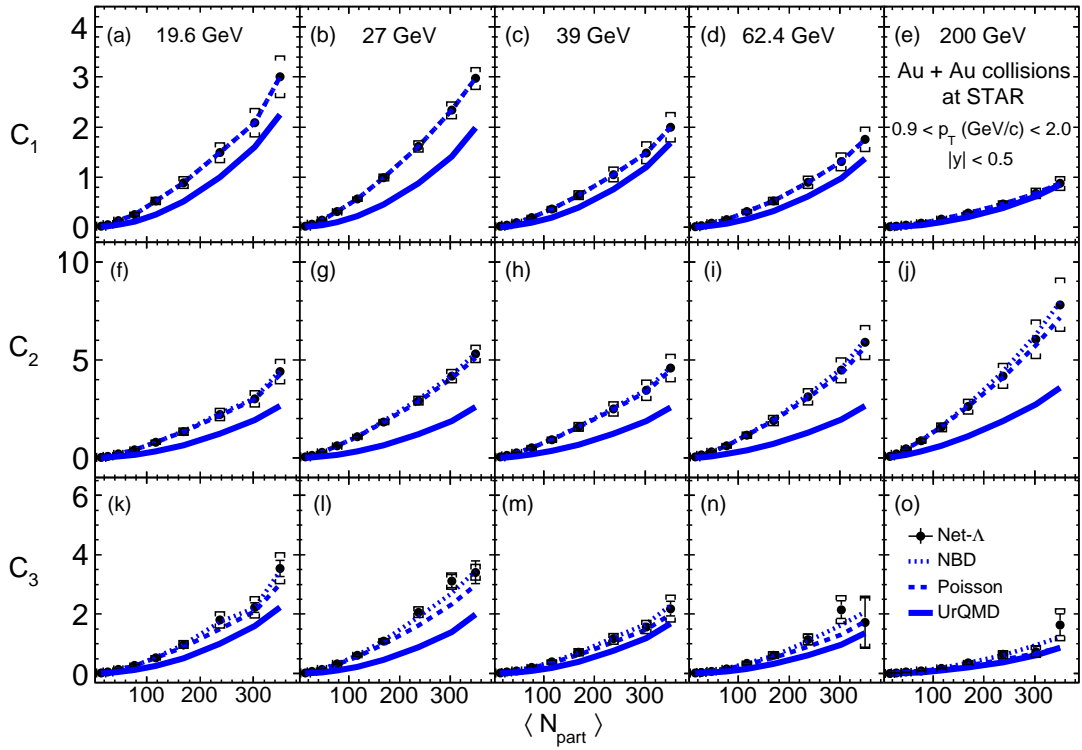}
\caption{\label{F_1}Centrality dependence of first three single cumulants $C_{1}$, $C_{2}$ and $C_{3}$ of net-$\Lambda$ multiplicity distributions at Au + Au collision energies $\sqrt{s_{NN}}$ = 19.6, 27, 39, 62.4 and 200 GeV. NBD and Poisson baselines are presented by dashed lines. UrQMD predictions are shown in solid lines. Black vertical lines represent the statistical uncertainties and caps represent systematic uncertainties. Results are corrected for the reconstruction efficiency and the CBWC is applied.}
\end{figure*}

\noindent{where $a$ and $b$ stand for the lower and upper bounds of the corresponding $p_{T}$ bin. $MC$ and $RC$ represent the tracks generated using Monte Carlo and the tracks reconstructed, respectively. The statistical uncertainty in the $V^{0}$ reconstruction efficiency was estimated, by following the procedure in Ref.\cite{X_Zu}, to be 2.25\% of the numbers quoted in Table II. Within the uncertainties, $\Lambda$ and $\overline{\Lambda}$ reconstruction efficiencies show negligible differences and are treated equally in the efficiency correction.}

\par

The collision centrality determination was done using the efficiency uncorrected number of identified charged particles (also known as reference multiplicity) in the pseudo-rapidity interval $|\eta|$ $<$ 1.0 by excluding protons and anti-protons in order to minimize possible self-correlations \cite{Xiaofeng:2013:AutoCorr}. The classification of events into different collision centrality classes was achieved by using this reference multiplicity along with the Glauber model \cite{Miller:2007:Glauber} simulations, in accordance with the procedure used in the published STAR net-proton analysis \cite{Net_P_STAR}. With this definition, the analysis was performed in nine collision centrality classes: 0-5\%, 5-10\%, 10-20\%, 20-30\%, 30-40\%, 40-50\%, 50-60\%, 60-70\%, and 70-80\%. 

\par

The reconstruction efficiency depends on the collision centrality. Therefore, the efficiency correction was applied in each centrality class separately. As in previous STAR analyses, a binomial distribution of the efficiency loss was assumed \cite{Luo:2018ofd}.  Volume fluctuations due to the selection of finite centrality bins can lead to the so-called finite bin width effect. The results presented here were corrected for this effect by applying the centrality bin width correction (CBWC) \cite{Xiaofeng:2013:AutoCorr}. 

\par

For the net-$\Lambda$ cumulants up to the  $3^{rd}$ order, the statistical uncertainties estimated using the delta theorem method \cite{Xiaofeng:2013:AutoCorr,Xiaofeng:2011tp:StatErr,Xiaofeng:2015:StatErr} and sub-sampling method gave similar values when the number of sub samples is greater than 10. The statistical uncertainties presented in this paper were estimated using the sub-sampling method with more than 10 sub-samples in all collision energies.

\par

The estimation of systematic uncertainties was done by varying the following topological and track cuts: 1.) simultaneous variation of distance of closest approach ($DCA$) of $p$($\overline{p}$) to primary vertex ($PV$) and $DCA$ of $\pi^{-}$ ($\pi^{+}$) to $PV$, 2.) $DCA$ of $p$($\overline{p}$) to $\pi^{-}$ ($\pi^{+}$), and 3.) simultaneous variation of $n\sigma_{p(\overline{p})}$ and $n\sigma_{\pi^{-}(\pi^{+})}$. In addition, variations of the estimated uncertainty on the reconstruction efficiency ($\pm$ 2.25\%) were also included in the systematic uncertainty estimation. All sources were treated as uncorrelated. Table III shows a detailed breakdown of all relevant contributions at the highest beam energy. The variations as a function of the collision energy are small, and the typical systematic uncertainties are on the order of 15\% for $C_{1}$, 18\% for $C_{2}$ and 30\% for $C_{3}$. The biggest contribution comes from the uncertainty in the $dE/dx$ measurements for particle identification ($n\sigma$ variation). The statistical uncertainties are presented by black vertical bars, and systematic uncertainties are presented by black caps in the figures of this paper. 

\begin{table}[hbt]
    \caption{Systematic uncertainty contributions from different sources (cuts and efficiency variations), $n\sigma_{p}$ and $n\sigma_{\pi}$ (NS), DCA of $V^{0}$ to $p$ and DCA of $V^{0}$ to $\pi$ (DCA-1), DCA of $p$ to $\pi$ (DCA-2) and efficiency variation (EV) for $\sqrt{s_{NN}} = 200 GeV$ collisions in 0-5\% centrality.}
    \medskip
    \centering
    \begin{tabular}{c | c | c | c | c | c}
    \hline\hline
    \backslashbox{Cumulant}{Source} & NS & DCA-1 & DCA-2 & EV & Total\\
    \hline\hline
    $C_{1}$ & 10.8\% & 2.3\% & 0.6\% & 1.6\% & 15.3\%\\
    $C_{2}$ & 11.7\% & 2.7\% & 1\% & 1.8\% & 17.2\%\\
    $C_{3}$ & 23.1\% & 4.4\% & 2.6\% & 2.8\% & 32.9\%\\
    $C_{2}/C_{1}$ & 1.3\% & 0.6\% & 0.4\% & 0.1\% & 2.4\%\\
    $C_{3}/C_{2}$ & 14.4\% & 5.5\% & 1.8\% & 1\% & 22.7\%\\
    \hline\hline
    \end{tabular}
    \label{tab:my_label_03}
\end{table}

\section{\label{sec:level2}Results}

\begin{figure*}
\centering
\includegraphics[width=0.7\textwidth]{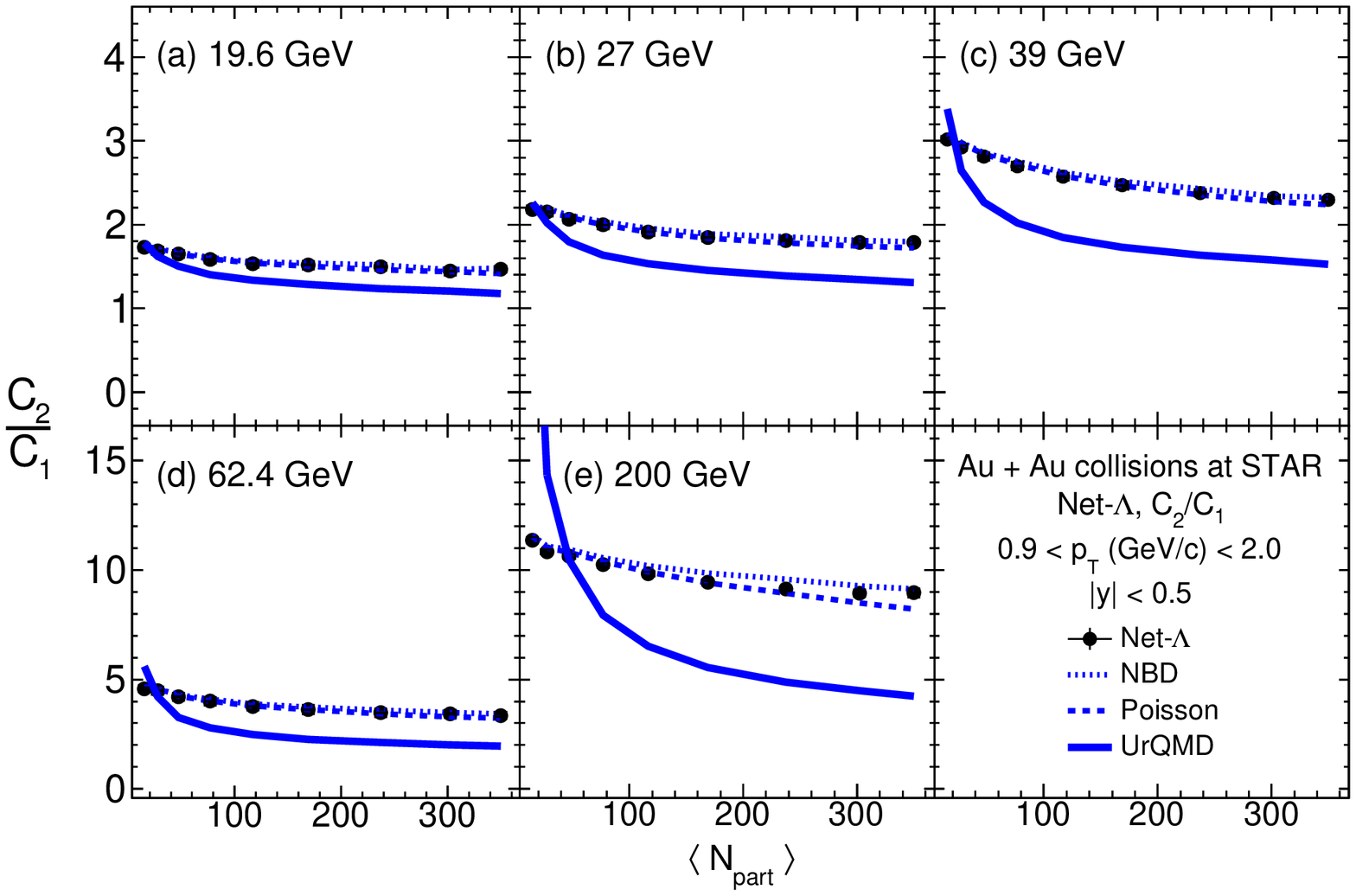}
\caption{\label{F_2}Centrality dependence of net-$\Lambda$ cumulant ratio, $C_{2}$/$C_{1}$ at Au + Au collision energies $\sqrt{s_{NN}}$ = 19.6, 27, 39, 62.4 and 200 GeV. NBD and Poisson baselines are presented by dashed lines. UrQMD predictions are shown in solid lines. Black vertical lines represent the statistical uncertainties and caps represent systematic uncertainties. Results are corrected for the reconstruction efficiency, and the CBWC is applied.}
\end{figure*}

\begin{figure*}
\centering
\includegraphics[width=0.7\textwidth]{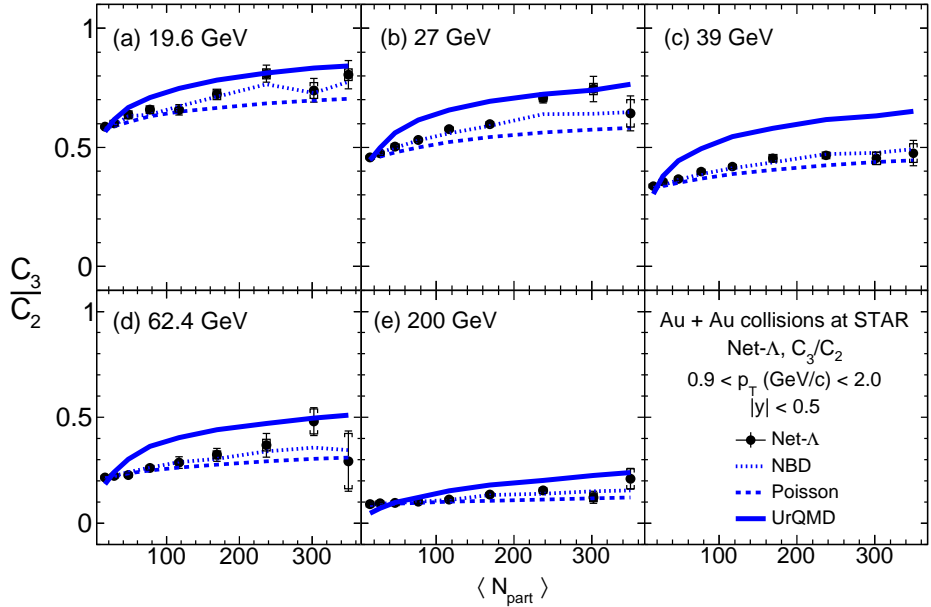}
\caption{\label{F_3}Centrality dependence of net-$\Lambda$ cumulant ratio, $C_{3}$/$C_{2}$ at Au + Au collision energies $\sqrt{s_{NN}}$ = 19.6, 27, 39, 62.4 and 200 GeV. NBD and Poisson baselines are presented by dashed lines. UrQMD predictions are shown in solid lines. Black vertical lines represent the statistical uncertainties and caps represent systematic uncertainties. Results are corrected for the reconstruction efficiency, and the CBWC is applied.}
\end{figure*}

The centrality dependence of the net-$\Lambda$ single cumulants ($C_{1}$, $C_{2}$ and $C_{3}$) in Au + Au collisions is presented in Fig. 2 for five beam energies from $\sqrt{s_{NN}}$ = 19.6 GeV to 200 GeV. The collision centrality is represented by the average number of participant nucleons ($\langle N_{\rm part} \rangle$) obtained from the Glauber model simulations. All three single cumulants show a steady increase as a function of increasing collision centrality at all collision energies due to the system volume dependence. For a fixed centrality, odd cumulants decrease as a function of increasing collision energy, which indicates that the $\Lambda/\overline{\Lambda}$ ratio approaches unity at the highest RHIC energies. The Poisson baselines were calculated by considering the means of individual particle distributions ($\Lambda$ and $\overline{\Lambda}$), while the NBD expectations were calculated using both means and variances. The Poisson and NBD expectations both show agreement with the measured single cumulants at all energies, within the statistical and systematic uncertainties. Odd cumulants of net-$\Lambda$ distributions are somewhat described by UrQMD predictions, in particular at higher energies, while there is a more significant disagreement at all energies with the measured $C_{2}$.

\par

The volume independent cumulant ratios, $C_{2}/C_{1}$ ($=$ $\sigma^{2}/M$) and $C_{3}/C_{2}$ ($=$ $S\sigma$), of net-$\Lambda$ distributions are presented in Figs. 3 and 4, respectively, as a function of collision centrality in Au + Au collisions for five beam energies from $\sqrt{s_{NN}}$ = 19.6 GeV to 200 GeV. The net-$\Lambda$ $C_{2}/C_{1}$ is nearly independent of the collision centrality due to the volume independence of the cumulant ratios, while it increases as a function of collision energy for a given centrality class, which is dominated by the energy dependence of $C_{1}$. Both the Poisson and NBD expectations show good agreement with the data. However, the UrQMD predictions show deviations from the data, which increase as a function of increasing collision energy. The major contribution to these deviations comes from the predictions for the net-$\Lambda$ $C_{2}$. The net-$\Lambda$ $C_{3}/C_{2}$ measurement also shows only a weak dependence on the collision centrality. It decreases as a function of increasing collision energy mainly due to the energy dependence of the net-$\Lambda$ $C_{3}$. The net-$\Lambda$ $C_{3}/C_{2}$ shows better agreement with the NBD expectations than with the Poisson baseline within the uncertainties, which could be an indication of less intra-event correlations between the produced $\Lambda$'s and $\overline{\Lambda}$'s \cite{Tarnowsky:2012vu:NBD}.

\begin{figure}[hbt]
\centering
\includegraphics[width=0.45\textwidth]{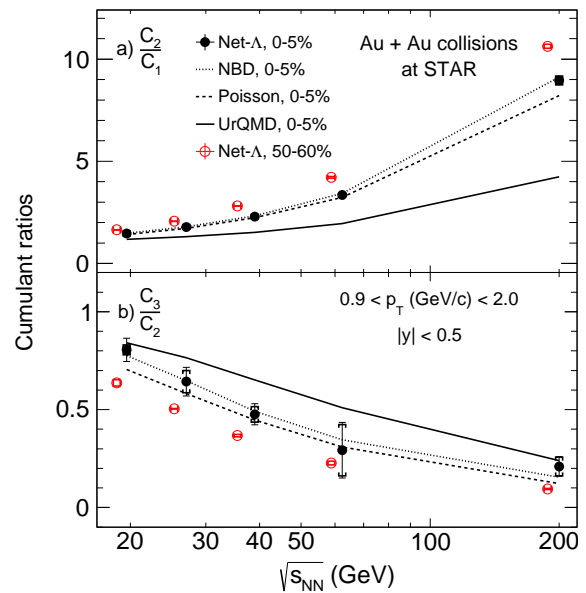}
\caption{\label{F_4}Beam energy dependence of  net-$\Lambda$ cumulant ratios, $C_{2}/C_{1}$ and $C_{3}/C_{2}$ in most central (0-5\%) and peripheral (50-60\%) Au + Au collisions. NBD and Poisson baselines are presented by dashed lines. UrQMD predictions are shown in solid lines. Black vertical lines represent the statistical uncertainties and caps represent systematic uncertainties. Results are corrected for the reconstruction efficiency and the CBWC is applied. The red data points are shifted left for clarity.}
\end{figure}

\begin{figure*}[hbt]
\centering
\includegraphics[width=0.85\textwidth]{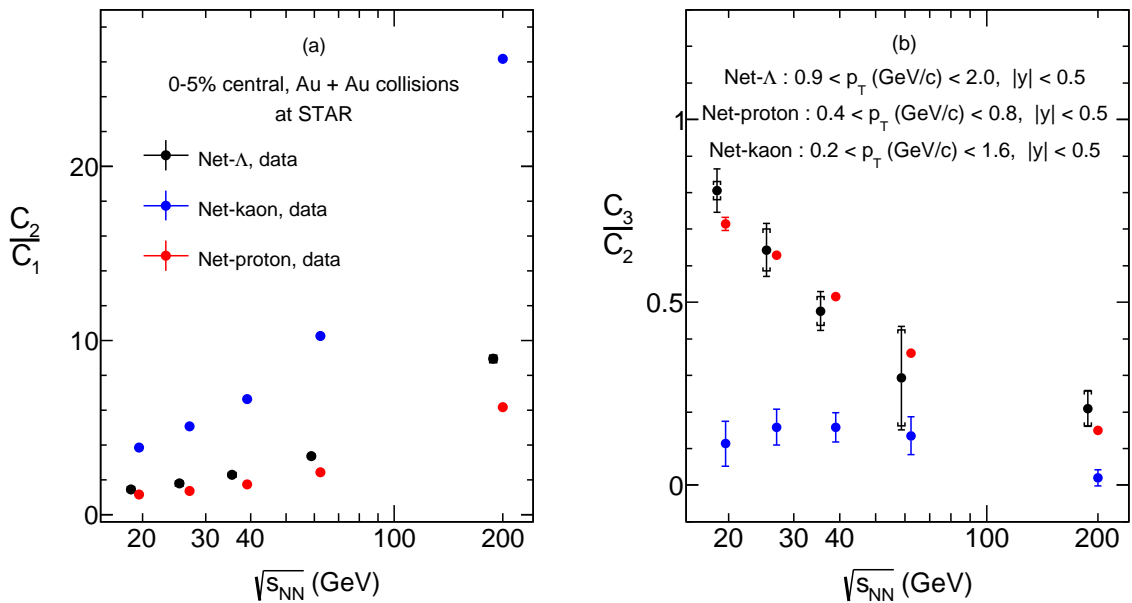}
\caption{\label{F_E1}Beam energy dependence of  net-$\Lambda$ (in black), net-proton (in red) and net-kaon (in blue) cumulant ratios, (a) $C_{2}/C_{1}$ and (b) $C_{3}/C_{2}$ in most central (0-5\%) Au + Au collisions. The vertical lines represent the statistical uncertainties and caps represent systematic uncertainties. Results are corrected for the reconstruction efficiency and the CBWC is applied. The black data points are shifted left for clarity.}
\end{figure*}

\begin{figure}[h!]
\centering
\includegraphics[width=0.47\textwidth]{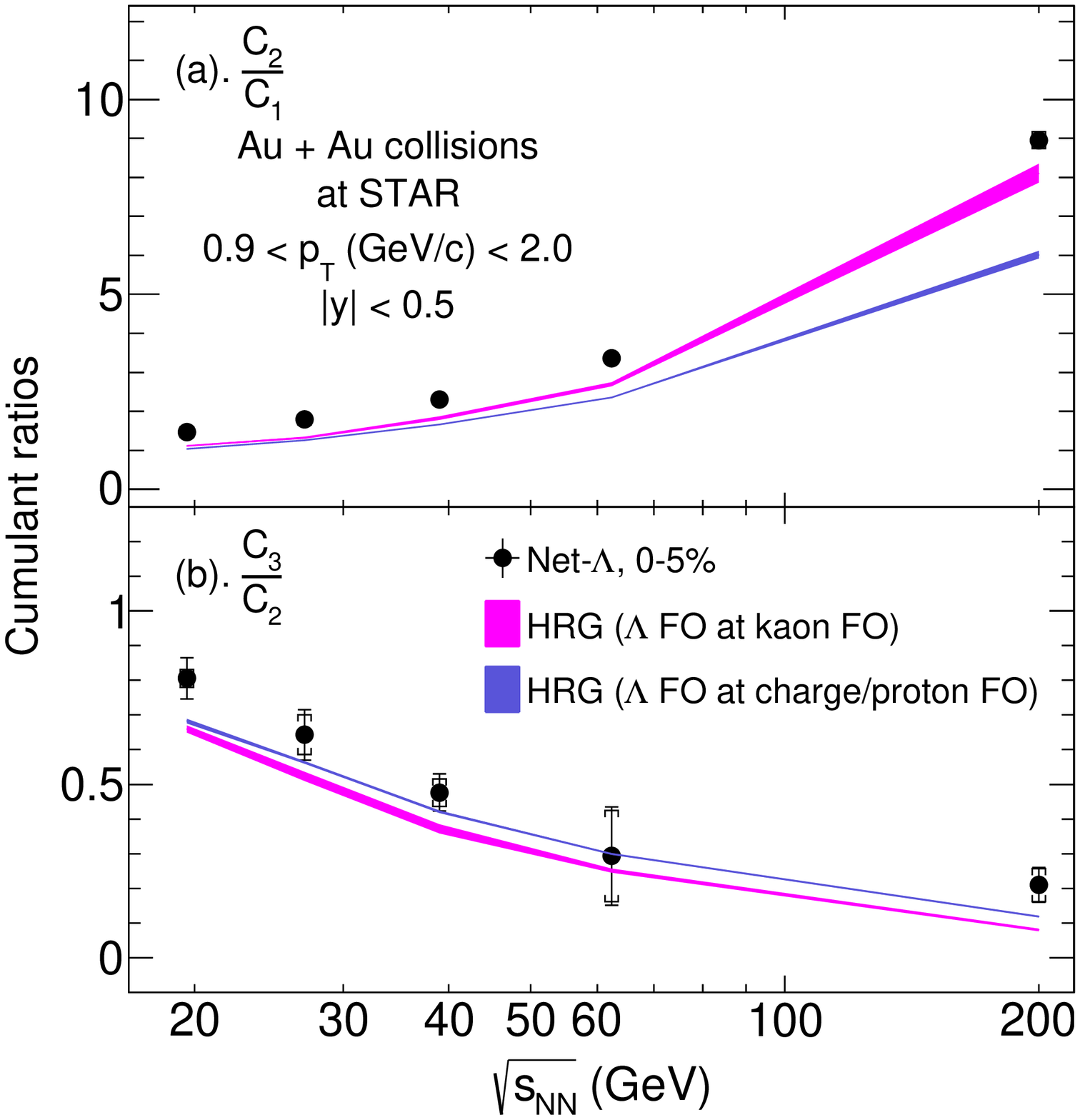}
\caption{\label{F_5}Black markers show the beam energy dependence of the measured net-$\Lambda$ cumulant ratios, (a) $C_{2}/C_{1}$ and (b) $C_{3}/C_{2}$ in most central (0-5\%) Au + Au collisions. Magenta bands show the net-$\Lambda$ cumulant ratios from a HRG calculation \cite{Bellwied:2018tkc} assuming $\Lambda$'s freeze-out under the same FO conditions as the kaons. Blue bands show the net-$\Lambda$ cumulant ratios from the same HRG calculation assuming $\Lambda$'s freeze out under the same FO conditions as the charged particles/protons. Results are corrected for the reconstruction efficiency, and CBWC is applied. The vertical bars and the caps represent the statistical and systematic uncertainties, respectively. Uncertainties in the HRG calculations are shown by the width of the bands.}
\end{figure}

\par

The energy dependence of net-$\Lambda$ $C_{2}/C_{1}$ and $C_{3}/C_{2}$ in the most central (0-5\%) and the peripheral (50-60\%) Au + Au collisions for the five beam energies is presented in Fig. 5. The Poisson, NBD and UrQMD expectations are shown for cumulant ratios measured in 0-5\% central collisions. Both the Poisson and NBD expectations show agreement with the measured net-$\Lambda$ $C_{2}/C_{1}$ and $C_{3}/C_{2}$ within statistical and systematic uncertainties, except at 200 GeV where the data are better described by the NBD expectations than by the Poisson baseline. UrQMD significantly deviates from the measured net-$\Lambda$ $C_{2}/C_{1}$ at all energies above 19.6 GeV. The net-$\Lambda$ $C_{3}/C_{2}$ measured at $\sqrt{s_{NN}} = 19.6$ and $200$ GeV agrees with the UrQMD predictions for the most central collisions, while it deviates from UrQMD expectations significantly at $\sqrt{s_{NN}} =  27, 39$ and $62.4$ GeV energies. There is no discernible presence of non-monotonic behavior, that could indicate critical fluctuations, in the measured net-$\Lambda$ cumulant ratios as a function of collision energy.

Figure 6 compares the ratio measurements obtained for net-$\Lambda$ to the ones for net-kaons \cite{Net_K_STAR} and net-protons \cite{Net_P_STAR}. The net-$\Lambda$ data follow more closely the net-proton data, which can be understood since the abundance and imbalance between particle and anti-particle for the $\Lambda$ baryon is more closely aligned with the proton numbers than with the mesonic strange state \cite{DDD, EEE}.

\begin{figure}[h]
\centering
\includegraphics[width=0.46\textwidth]{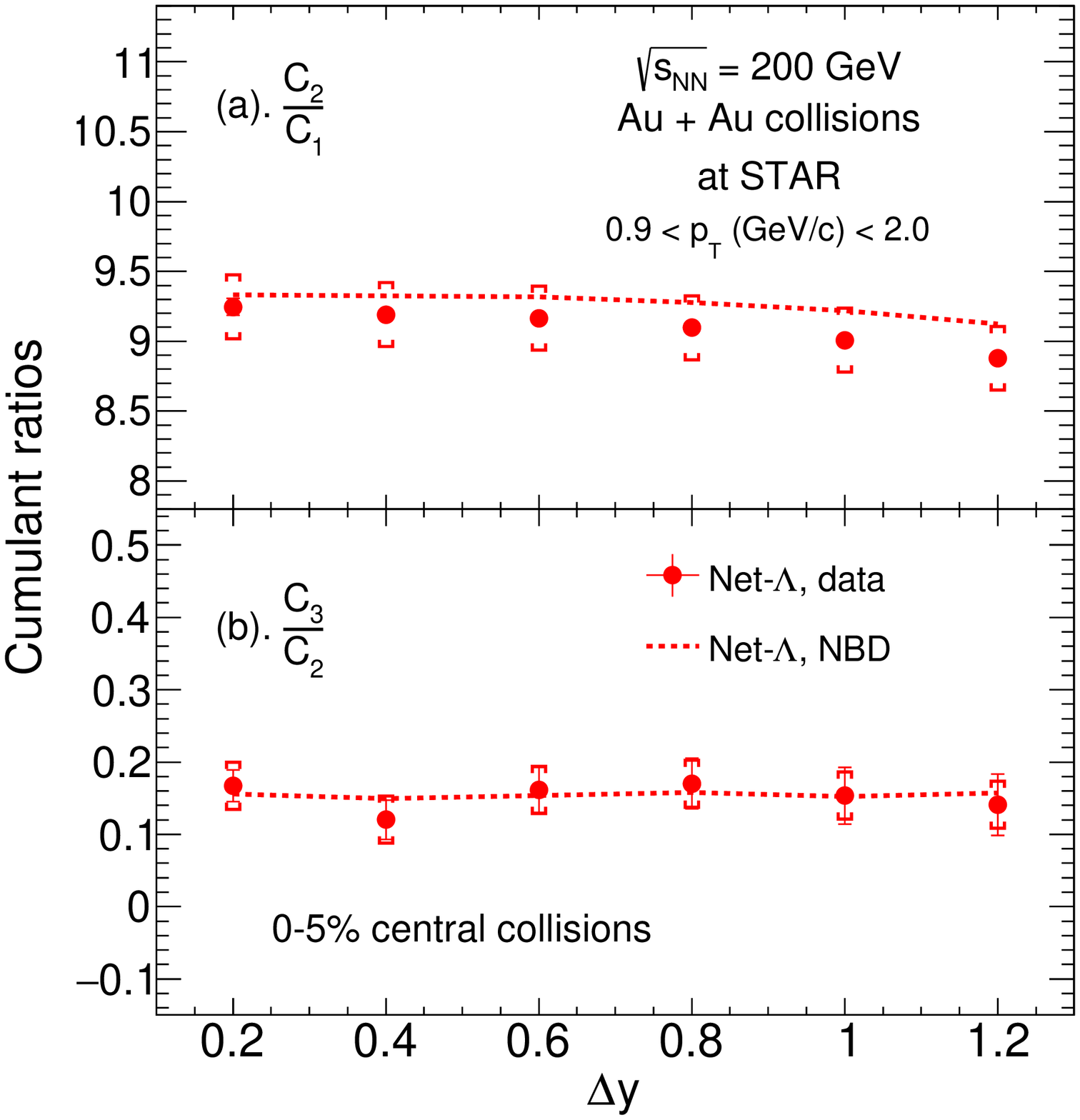}
\caption{\label{F_6}Rapidity dependence of net-$\Lambda$ cumulant ratios, (a) $C_{2}/C_{1}$ and (b) $C_{3}/C_{2}$ in $\sqrt{s_{NN}} = 200$ GeV Au+Au collisions at 0-5\% centrality. Dashed lines show the NBD expectations. Vertical error bars represent the statistical uncertainties and caps represent the systematic uncertainties. Results are corrected for the reconstruction efficiency and CBWC is applied.}
\end{figure}

\begin{figure}[h]
\centering
\includegraphics[width=0.52\textwidth]{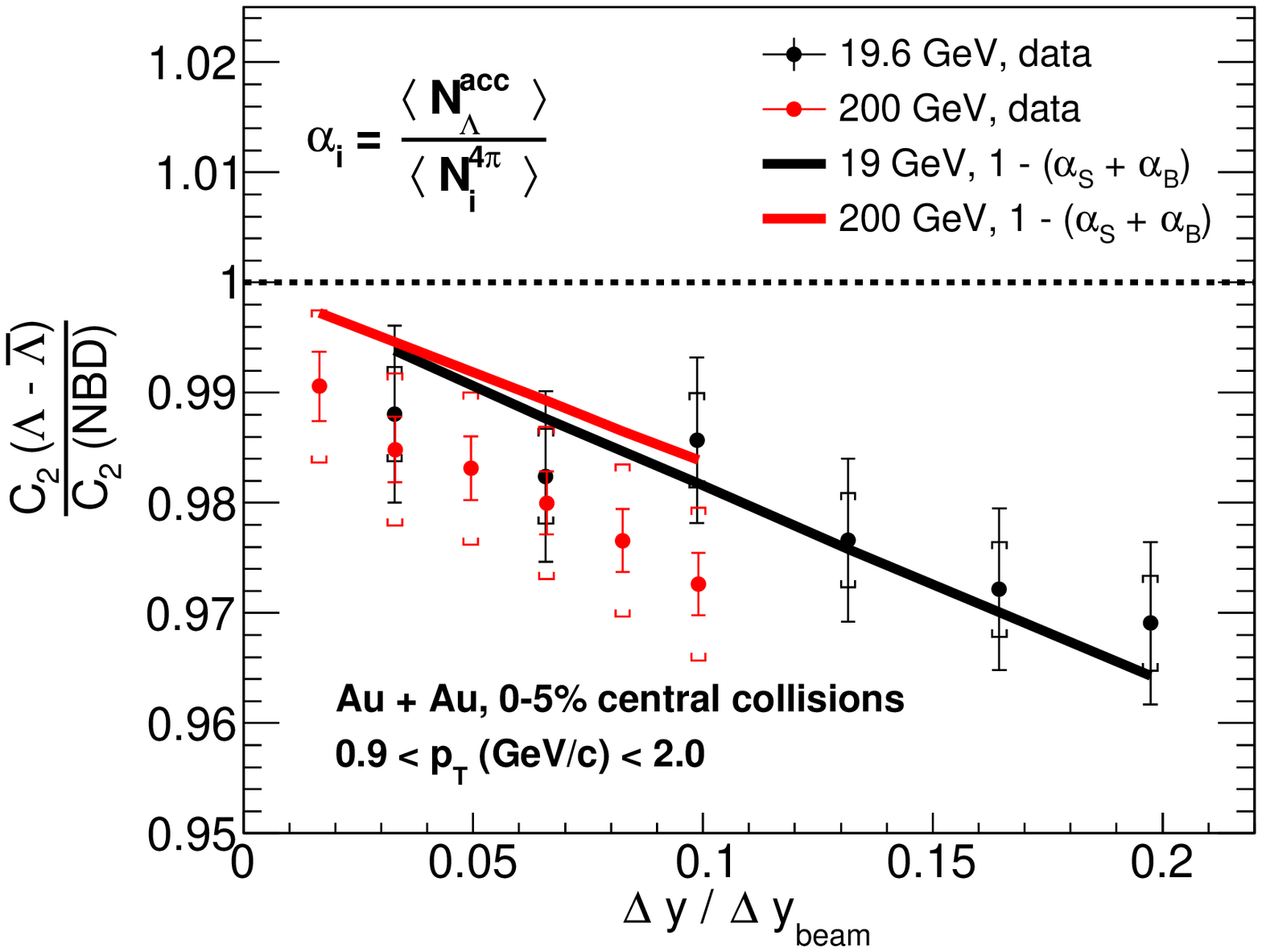}
\caption{\label{F_7} Rapidity dependence of the normalized $C_{2}$($\Lambda$-$\overline{\Lambda}$) in most central (0-5\%) collisions for 19.6 and 200 GeV collision energies. The solid lines show the expected effects from baryon number (B) and strangeness (s) conservation.}
\end{figure}

\par

In order to determine chemical freeze-out parameters, the energy dependence of the lowest net-$\Lambda$ cumulant ratio ($C_{2}/C_{1}$) in most central Au + Au collisions is compared to the cumulant ratios calculated, assuming different freeze-out (FO) conditions using the latest HRG model \cite{Bellwied:2018tkc}. As shown in \cite{Koch_Fluc_Corr}, ratios that contain higher order moments are more sensitive to dynamical effects and thus lead to more unreliable results when FO parameters are extracted.  We therefore focused on the high resolution $C_{2}/C_{1}$ measurement for our conclusions. Previous HRG model parameters that were based on the STAR net-proton/net-charge measurements \cite{Alba:2014} and net-kaon measurements \cite{Bellwied:2018tkc}, were used as benchmarks to compare to the net-$\Lambda$ measurements. In these earlier calculations a difference in freeze-out temperature of about 20 MeV was obtained between strange and light quark particles.

From the comparison of measured net-$\Lambda$ $C_{2}/C_{1}$ ratios with HRG predictions in Fig. 7(a), it is apparent that the measured net-$\Lambda$ $C_{2}/C_{1}$ ratio is closer to the $C_{2}/C_{1}$ ratio calculated assuming the kaon freeze-out conditions than the proton/charge freeze-out conditions. In other words the difference between the HRG calculations to the measured net-$\Lambda$ $C_{2}/C_{1}$ ratio becomes small when the freeze-out conditions extracted from the net-kaon fluctuations were used in the prediction. This observation is not trivial to interpret, but one possible explanation is that the strangeness conservation  plays a more prominent role for the $\Lambda$ baryon at freeze-out than the baryon number. For a complete understanding of this observation, further investigations, both theoretically and experimentally, are necessary. It is important to note that the measured net-$\Lambda$ $C_{2}/C_{1}$ ratio is much closer to the measured net-proton $C_{2}/C_{1}$ ratio (see Fig. 6(b)), whereas the deduced freeze-out temperature from the net-$\Lambda$ $C_{2}/C_{1}$ ratio is closer to the deduced temperature from the net-kaon $C_{2}/C_{1}$ ratio (see Fig.7(a)). This points at the fact that the measured net-cumulants are dominated by the particle to anti-particle ratios, but the deduced temperatures are driven by the resonance contributions to the final ratios.  

In contrast to any critical endpoint searches, the chemical freeze-out is presently best determined in the lower cumulant ratios that have the smallest error bars. The temperature differences found in the HRG calculations are less pronounced in the $C_{3}/C_{2}$ ratio shown in Fig.7(b) and the error bars of the measurement are larger, which prohibits a definitive statement on the basis of the higher cumulant ratio.

\par

Finally, the rapidity dependence of net-$\Lambda$ cumulant ratios has been investigated for the most central 200 GeV Au + Au collisions. Figure 8 shows a comparison of the cumulant ratios to NBD expectations as a function of $\Delta y$, i.e. the rapidity window around mid-rapidity. Both cumulant ratios show a weak dependence on the selected rapidity window and generally good agreement with the NBD baseline. Figure 9 presents the ratio of the net-$\Lambda$ $C_2$ to a NBD baseline as a function of relative rapidity coverage $\Delta y$ normalized by $\Delta y_{\rm beam}$, for $\sqrt{s_{NN}} = 19.6$ and 200 GeV Au+Au collisions at 0-5\% centrality.

The potential impact of baryon number conservation as a function of the detector acceptance has been addressed recently in the literature \cite{Braun-Munzinger_rapidity, FFF}. When comparing our results to a calculation based on this approach, but taking into account the fact that the $\Lambda$ carries both baryon number and strangeness, we find the solid lines shown in Fig.9. Here the acceptance factors ($\alpha_B$ and $\alpha_s$) are defined as the ratio between the average number of $\Lambda$'s in the selected acceptance ($\langle N^{\rm acc}_{\Lambda} \rangle$) over the average number of total baryons  ($\langle N^{4\pi}_{B} \rangle$) or total strangeness ($\langle N^{4\pi}_{s} \rangle$) in the full phase space, respectively. $\langle N^{4\pi}_{s} \rangle$ is defined as the number of strange quarks confined in strange hadrons in 4$\pi$. The quantities $\langle N^{4\pi}_{B} \rangle$ and $\langle N^{4\pi}_{s} \rangle$ were calculated using the UrQMD model. The 4$\pi$ particle yields are adequately described by this model in the BES energy range. The acceptance factors, $\alpha_B$ and $\alpha_s$, were calculated separately and added to approximate the anticipated correction due to the combined quantum number conservation effect. The relative contribution of each acceptance factor is slightly energy dependent. At the lower energy $\alpha_s$ dominates, whereas at the highest RHIC energies the $\alpha_B$ contribution is stronger. As shown in Fig. 9, the combined factor accounts well for the rapidity dependence at $\sqrt{s_{NN}} = 19.6$ GeV. The beam energy dependence is not perfectly captured, but one should note that the overall effect on the final result is small at all energies ($<$3\%), which reflects the number of protons and $\Lambda$'s in the detector acceptance. A more detailed study of the conservation effects in the Beam Energy Scan regime, i.e. in the region where the varying baryon stopping plays an important role, might also require a more local baryon number conservation approach, as was suggested recently \cite{Pruneau:2019baa}.

\section{\label{sec:level4}Summary}

\par

The fluctuations of conserved quantum numbers such as net-charge, net-baryon number and net-strangeness provide useful information about the nuclear matter phase transition. Fluctuations of measured net-particle multiplicity distributions in the medium produced after the heavy-ion collisions have been successfully used as proxies for conserved quantum number fluctuations. The measured cumulant ratios of net-charge, net-proton and net-kaon multiplicity distributions at STAR have been compared to HRG calculations to extract the freeze-out parameters in the QCD crossover region. The $\Lambda$ baryon carries both baryon and strangeness quantum numbers. In this paper, the first measurements of the centrality, collision energy and rapidity dependence of net-$\Lambda$ single cumulants ($C_{1}$, $C_{2}$, $C_{3}$) and cumulants ratios ($C_{2}/C_{1}$, $C_{3}/C_{2}$) in Au + Au collisions measured at the STAR detector for five beam energies ($\sqrt{s_{NN}}$ = 19.6, 27, 39, 62.4 and 200 GeV) were presented. Results were compared to Poisson, NBD, UrQMD and HRG model expectations. The comparison of these net-$\Lambda$ fluctuation measurements to the latest HRG model provides important information needed in understanding the dominant quantum number specific contributions to the hadronization process. 

\par

Both Poisson and NBD expectations were able to describe the centrality and collision energy dependence of the net-$\Lambda$ single cumulants ($C_{1}$, $C_{2}$ and $C_{3}$) and cumulant ratios ($C_{2}$/$C_{1}$ and $C_{3}$/$C{2}$) fairly well. This indicates that any onset of critical or non-monotonic behavior is not apparent for the energies and cumulants studied here, which is consistent with the results of previous STAR/BES-I fluctuation analyses.

\par

In $\sqrt{s_{NN}} = 200$ GeV Au+Au collisions, the net-$\Lambda$ cumulant ratios show better agreement with the NBD expectations than with the Poisson baseline. The UrQMD predictions show good agreement with the odd cumulants within the uncertainties at 200 GeV, while there is a considerable deviation from the measured $C_{2}$ at all energies. The rapidity dependence of the net-$\Lambda$ cumulant ratios, based on our results at 200 GeV, is small. 
Nevertheless, single cumulants deviate at the level of a few percent from the NBD baseline over the measured rapidity interval at all beam energies, which cannot be attributed to baryon number conservation alone. Only when baryon number and strangeness conservation are combined is the difference well explained. 

The lowest net-$\Lambda$ cumulant ratio, $C_{2}/C_{1}$, was compared to recent HRG model calculations in order to obtain the proper chemical freeze-out scenario for the strange baryon. The deviation of the HRG calculations from the measured $C_{2}/C_{1}$ ratio becomes small when the HRG model calculations were performed using the freeze-out conditions extracted considering strange (kaon) net-particle fluctuations, but not when the non-strange (charge/proton) net-particle fluctuations were considered. Further experimental and theoretical progress is needed in order to understand whether this observation is due to a flavor hierarchy in the freeze-out parameters, which could lead to possible sequential hadronization.   

\section{\label{sec:level5}Acknowledgement}

We thank the RHIC Operations Group and RCF at BNL, the NERSC Center at LBNL, and the Open Science Grid consortium for providing resources and support.  This work was supported in part by the Office of Nuclear Physics within the U.S. DOE Office of Science, the U.S. National Science Foundation, the Ministry of Education and Science of the Russian Federation, National Natural Science Foundation of China, Chinese Academy of Science, the Ministry of Science and Technology of China and the Chinese Ministry of Education, the National Research Foundation of Korea, Czech Science Foundation and Ministry of Education, Youth and Sports of the Czech Republic, Hungarian National Research, Development and Innovation Office, New National Excellency Programme of the Hungarian Ministry of Human Capacities, Department of Atomic Energy and Department of Science and Technology of the Government of India, the National Science Centre of Poland, the Ministry  of Science, Education and Sports of the Republic of Croatia, RosAtom of Russia and German Bundesministerium fur Bildung, Wissenschaft, Forschung and Technologie (BMBF) and the Helmholtz Association.

\bibliographystyle{apsrev4-1}
\bibliography{main.bib}

\end{document}